\documentclass[mathleft,fleqn,%
]{an} 
\usepackage{longtable}
\usepackage{tablefootnote}
\usepackage{graphicx}
\usepackage[varg]{txfonts}
\usepackage{apacite}
\usepackage[authoryear]{natbib}
\begin{document}

\Pagespan{1}{}
\Yearpublication{2014}%
\Yearsubmission{2014}%
\Month{0}%
\Volume{999}%
\Issue{0}%
\DOI{asna.201400000}%

\title{Physical parameters of RR Lyrae stars in NGC 6171}

\author{D. Deras \inst{1}\fnmsep\thanks{Corresponding author:
        {dderas@astro.unam.mx}}
A. Arellano Ferro\inst{1},  S. Muneer\inst{2}, Sunetra Giridhar\inst{2}, \and R. Michel\inst{3}
}
\titlerunning{Physical parameters of RR Lyrae stars in NGC 6171}
\authorrunning{D. Deras et al.}
\institute{
Instituto de Astronom\'ia, Universidad Nacional Aut\'onoma de M\'exico, Ciudad
de M\'exico, CP 04510, M\'exico.
\and 
Indian Institute of Astrophysics, Koramangala 560034, Bangalore, India.
\and
Observatorio Astron\'omico Nacional, Instituto de Astronom\'ia, Universidad Nacional Aut\'onoma de M\'exico, Ap. P. 877, Ensenada, BC 22860, M\'exico.}

\received{XXXX}
\accepted{XXXX}
\publonline{XXXX}


\abstract{%
  We present an analysis of \emph{VI} CCD time-series photometry of the globular cluster NGC 6171. The main goal is to determine individual physical parameters for single-mode RR Lyrae stars present in the field of view of our images, by means of light curve Fourier decomposition and well-established calibrations and zero-points. This leads to the estimation of the mean values of the metallicity and distance for the parental cluster. From the RRab stars we find [Fe/H]$_{ZW}$ = -1.33 $\pm$ 0.12 and a distance $d$ = 5.3 $\pm$ 0.3 kpc and from the RRc stars we find [Fe/H]$_{ZW}$ = -1.02 $\pm$ 0.19 and a distance $d$ = 5.3 $\pm$ 0.2 kpc. Independent methods, such as the P-L relations for RRab and SX Phe stars enable the estimation of a weighted average distance to the cluster of 5.4 $\pm$ 0.1 kpc. We confirm the amplitude modulations of the Blazhko type of five RRab and the non-membership of V21 to the cluster. The colour-magnitude diagram is consistent with an age of 11 Gyrs. The distribution of RRab and RRc stars seems well segregated around the first overtone red edge of the instability strip. This positions NGC 6171 among OoI type clusters where the pulsating modes are neatly separated in the HB. We report two new irregular variables of the Lb type.} 
  
 \maketitle

\section{Introduction}

NGC 6171 (M107) is a very disperse globular cluster (apparent diameter $\sim 13'$) located in the constellation of Ophiuchus ($\alpha = 16^{h}32'31.86'', \delta = -13^{\circ} 03' 13.6''$, J2000), and very close to the Galactic bulge ($l$ =  $3.37^{\circ}$ , $b$ = $23.01^{\circ}$). In the compilation of \citet{Harris1996} (2010 edition), the distance from the Sun and the reddening for the cluster are given as 6.4 kpc and $E(B-V) = 0.33$ respectively.
The Catalog of Variable Stars in Globular Clusters (CVSGC; Clement et al. 2001; 2015 edition) lists 26 variables of which 23 are RR Lyrae, one Mira (V1), one Lb variable (V25) and one SX Phe (V26).  Oosterhoff discovered the first 24 variable stars \citep{Oo1938} while V25 was discovered by  \citet{Lloyd1973} and V26 by \cite{McCombs2012}.
The cluster membership of the RR Lyrae stars in NGC 6171 was determined by \citet{Cudworth1992}.\\ 

The historical data available for this cluster spans 82 years which enabled \citet{are2018} to perform an analysis of secular changes in the periods of the RR Lyrae stars present in the cluster. It was found that 82\% of the sample had stable periods and significant
period variations were found only in four stars, three of which had decreasing periods, and if interpreted as of evolutionary origin, it implies evolution towards the blue side of the Horizontal Branch (HB). However, no further analysis of the light curves has ever been performed. 

In the present paper, we report the analysis of time series photometry of NGC 6171 in the \emph{VI} bands. We made use of Fourier decomposition of the RR Lyrae stars light curves to find individual values of [Fe/H], distance, mass, effective temperature and radii and hence the average metallicity and distance to the parental cluster. We also briefly discuss the morphology of the Horizontal Branch and the age of the cluster. The paper is organised in the following way: in $\S$ 2 the observations and the transformation to the standard system are described\textbf{;} in $\S$ 3 we discuss the reddening of the cluster; in $\S$ 4 the Fourier light curve decomposition is performed and the physical parameters are presented, the Oosterhoff type and the period-amplitude diagram are discussed; in $\S$ 5 the distance to NGC 6171 derived from an assortment of methods is evaluated along with the implications of the adopted reddening; in $\S$ 6 the resulting iron abundance [Fe/H] is framed in the perspective of the numerous previous estimates; in $\S$ 7 the RR Lyrae distribution on the HB is highlighted along with overall consistency of the Colour-Magnitude Diagram (CMD) and the age of the cluster. Finally, in $\S$ 8 we summarise our conclusions.

\begin{figure} 
\includegraphics[width=8.0cm,height=5.0cm]{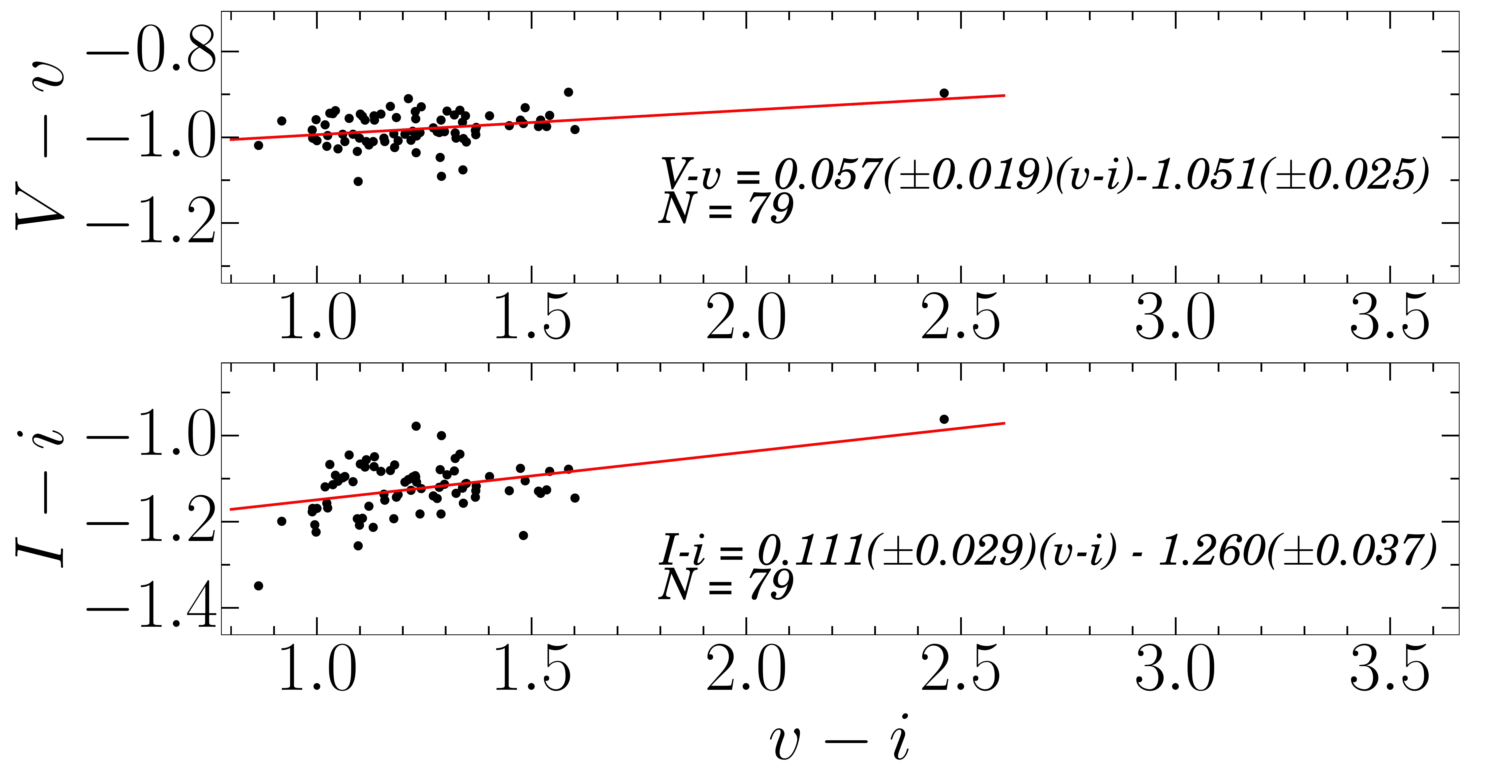}
\includegraphics[width=8.0cm,height=5.0cm]{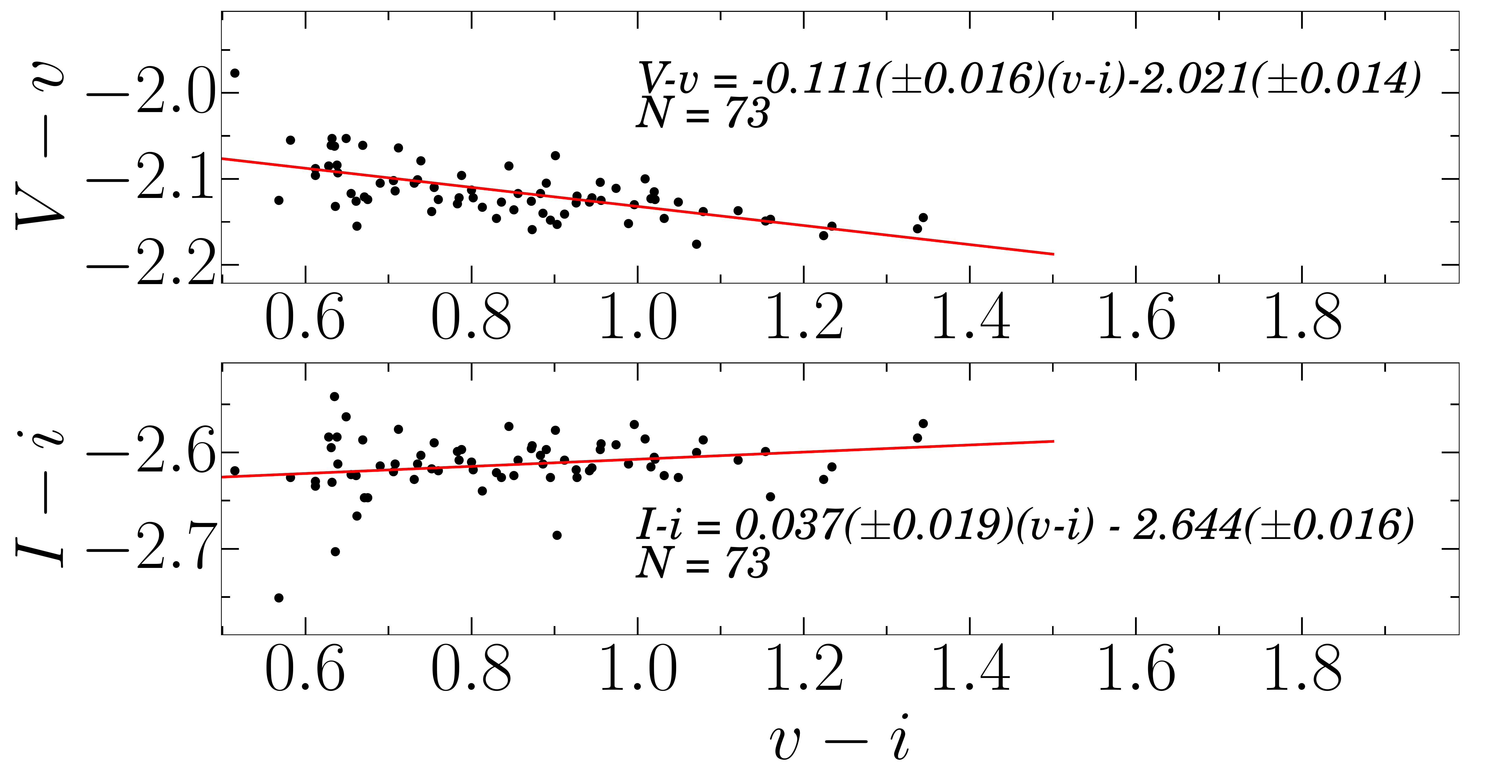}
\caption{Transformation equations in $V$ and $I$ filters between the instrumental and the standard photometric systems. These equations were calculated using a set of standard stars (N=79 for Hanle and N=73 for San Pedro M\'artir) in the field of NGC~6171. The top panel corresponds to Hanle and the bottom panel to San Pedro M\'artir observations.}  
    \label{transV}
\end{figure}

\section{Observations}

The observations used for the present work  were performed at two different locations. The first set of data was obtained using the 2m telescope of the Indian Astronomical Observatory (IAO) in Hanle, India on two epochs, each consisting of two consecutive nights. The first epoch spans the nights of June 26th-27th 2015, the second epoch spans the nights of May 18th-19th 2016. The detector used was a SITe002 CCD of 2048 $\times$ 2048 pixels with a scale of 0.296 arcsec/pix and a field of view (FoV) $\approx$ 10.1 arcmin$^{2}$. The second set of data was obtained with the 0.84m Ritchey-Chr\'etien telescope of the San Pedro M\'artir Observatory (SPM), M\'exico during 8 nights between June 28th to July 5th, 2017. The detector used was a 2048 $\times$ 2048 pixels ESOPO CCD (e2v CCD42-90) with 1.7 $e^{-}$/ADU gain, a readout noise of 3.8$e^{-}$, a 0.44361 arcsec/pix scale and a field of view $\approx$ 7.4 arcmin$^{2}$, along with the Mexman filter wheel. The log of the observations is given in Table
\ref{tab:observations}.\\

The observations described above were published in electronic format by \citet{are2018}. In that paper only the $V$ data of the RR Lyrae stars were employed to discuss the secular period variations. In the present paper we aim to calculate the physical parameters of RR Lyrae via the Fourier light curve decomposition and extend our analysis to the $V$ and $I$ bands for all variables present in the FoV of our images. In the following subsections we describe the transformation to the standard system.

\subsection{Difference image analysis}
We have employed Difference Imaging Analysis (DIA) technique with its 
pipeline implementation DanDIA (\citealt{Bramich2008}; \citealt{Bramich2013}) to obtain high-precision photometry for all the point sources in the FoV of the CCD for the reduction of our data. DanDIA creates a reference image by stacking the best quality images in each filter and then it subtracts them from the rest of the images in the collection. Each star's differential flux is then determined by means of the PSF calculated by DanDIA from a large number of isolated stars in the FoV, allowing for spatial variation on the chip. This enabled to construct an instrumental light curve for each star in our FoV. \\

\subsection{Transformation to the standard system}
\label{calibration}

Having obtained the instrumental light curves, we have applied the 
 methodology described by \cite{BF12} to correct for possible systematic photometric errors which tend to be significant for the brightest stars. The method solves for magnitude offsets that should be applied to each photometric measurement from the given image. In our case these offsets were of the order of 5 mag in the bright end and negligible for fainter stars.

The calibrated light curves of all the point sources in our FoV, were transformed to the Johnson-Kron-Cousins standard magnitude system \citep{Landolt1992}. We did this by using standard stars in our FoV identified in the catalogue of Photometric Standard Fields \citep{Stetson2000}. The transformation equations for each data set are shown in Fig. \ref{transV}, where the mild colour dependence of the standard minus the instrumental magnitudes is observed.

\begin{table}[t]
\footnotesize
\caption{Observations log of NGC~6171. Data are from two sites;
Hanle (Han) and San Pedro M\'artir (SPM).
Columns $N_{V}$ and $N_{I}$ give the number of images taken with the $V$ and $I$
filters respectively. Columns $t_{V}$ and $t_{I}$ provide the exposure time,
or range of exposure times. In the last column the average seeing is listed.}
\centering
\begin{tabular}{lcccccc}
\hline
Date  & Site &$N_{V}$ & $t_{V}$ (s) & $N_{I}$ &$t_{I}$ (s)&seeing (") \\
\hline
 20150626 & Han & 20 & 30-40 & 20 & 8     & 2.0 \\
 201505627 & Han & 22 & 30    & 22 & 8     & 2.1 \\
 20160518 & Han & 60 & 30    & 60 & 10	  & 2.2 \\
 20160519 & Han & 43 & 30    & 43 & 10    & 1.9  \\
 20170628 & SPM & 86 & 40-50 & 90 & 25-30 & 5.1 \\
 20170629 & SPM & 16 & 40    & 17 & 25    & 2.3 \\
 20170630 & SPM & 39 & 40    & 41 & 25    & 1.8 \\
 20170701 & SPM & 40 & 40    & 40 & 25    & 1.8 \\
 20170702 & SPM & 38 & 40    & 39 & 25    & 2.2 \\
 20170703 & SPM & 35 & 40    & 36 & 25    & 1.9 \\
 20170704 & SPM & 37 & 40    & 38 & 25    & 2.0 \\
 20170705 & SPM & 31 & 40    & 33 & 25    & 2.2 \\
\hline
Total:   &-- &467 &  -- & 479 & -- & --\\
\hline
\end{tabular}
\label{tab:observations}
\end{table}

\begin{figure*}[h]
   \centerline{\includegraphics[width=18cm, height=9.5cm]{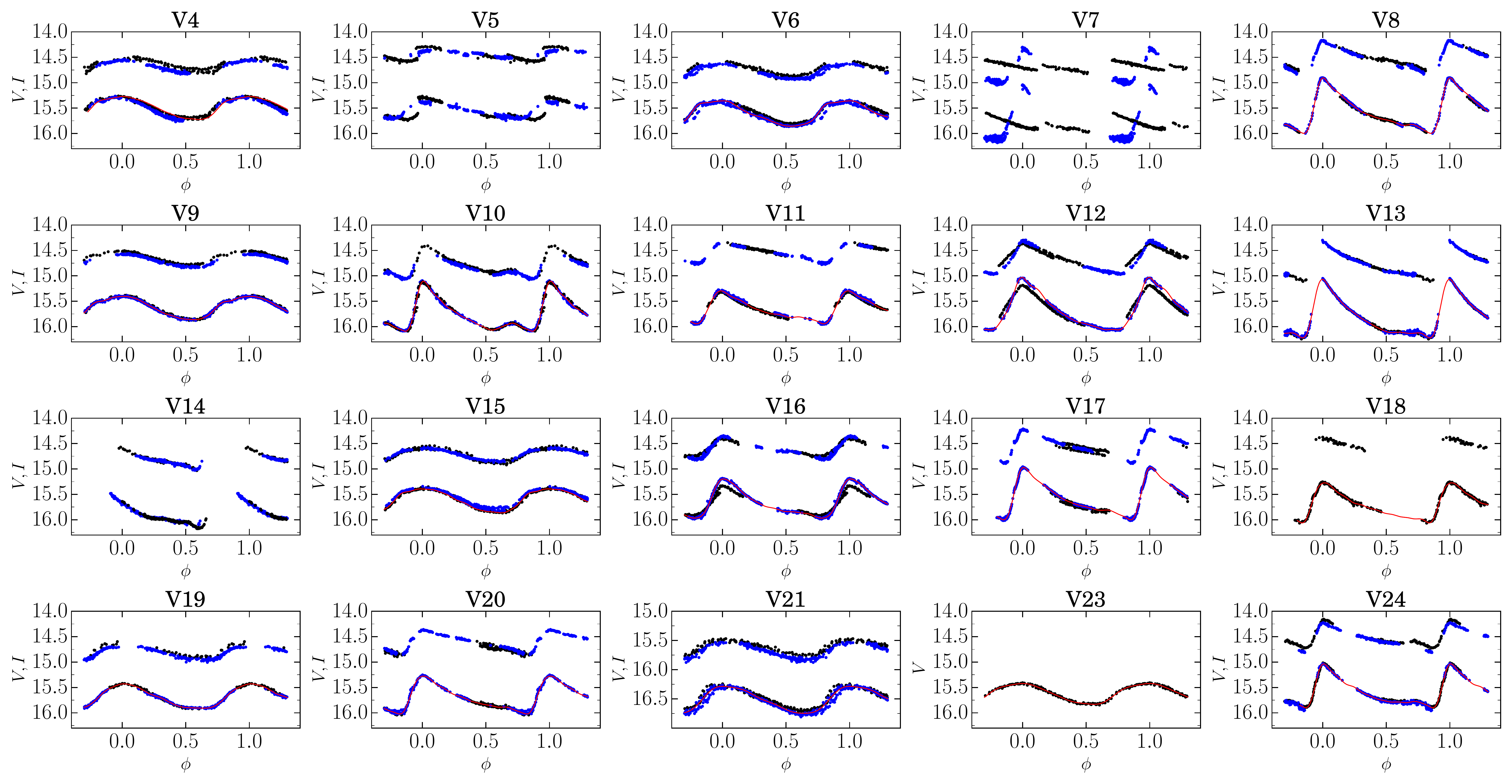}}
 \caption{Light curves in $V$ and $I$ filters of RR Lyrae stars in NGC 6171. The light curves are a combination of the data obtained at Hanle (black symbols) and at San Pedro M\'artir (blue symbols). The continuous red line represents the Fourier fit. The scale of the $V$-axis is the same for all plots with the exception of fainter V21 which is not a cluster member. In the case of V23, the $V$ light curve was taken from \citet{Clement1997}. The data for all this light curves have been published in electronic format by \cite{are2018}.}
\label{mosaico_6171}
\end{figure*}

\section{On the reddening of NGC 6171}
\label{reddening}

Earlier estimates of $E(B-V)$ in NGC 6171 span a large range, from 0.25 to 0.46, and were obtained by an assortment of methods (see Table II from \cite{Smith1986} for a summary) . For example, using a variety of reddening independent photometric colours favored values between 0.31 and 0.38 (\citealt{Zinn80}; \citealt{Zinn85}). The adopted value of $E(B-V)$ in the catalogue of \cite{Harris1996} is 0.33. The largest value of 0.46 has been estimated by \citet{vdb67}. 
Incompatibilities of the low reddening value with 100 $\mu$m dust emission data \citep{DuBi2000} and between photometric and spectroscopic $T_{\rm eff}$ for cluster giants, have been recently noted \citep{OConnell2011}. These authors point out that these inconsistencies can be alleviated if $E(B-V)$ is 0.45 and 0.46 respectively.

In an attempt to clarify the above incompatibilities we estimated the reddening for the RRab stars individually using the method explained by \cite{Sturch1966}, according to which these stars have a constant intrinsic color $(B-V)_o$ near minimum light, between phases 0.5 and 0.8. This, coupled with the calibration of $(V-I)_o$ in this range of phases  by \cite{Guldenschuh2005} as $\overline{(V-I)_{o,min}}$ = 0.58 $\pm$ 0.02 mag, allowed us to estimate the individual values of $E(V-I)$ for a sample of RRab stars. These were converted to $E(B-V)$ using the ratio $E(V-I)/E(B-V) = 1.259$ derived by \cite{Schlegel1998}. The light curves were phased with the periods listed in Table \ref{variables} and were adopted from the analysis of \citet{are2018}. The resulting values for the reddening are listed in the second-to-last column of Table \ref{tab:fourier_coeffs}. For V17 we have used minimum values of $V$ and $I$ to calculate $(V-I)_{min}$ as we were constrained by the data gap in 0.5 to 0.8 phases. Ignoring the value for V18, which has a prominent gap near minimum light, we derived an average $E(B-V)$ estimate of 0.45 $\pm$ 0.03. This value is in good agreement with those derived using reddening maps and the calibration  of \cite{Schlegel1998}, $0.46 \pm 0.01$. Although one could expect differential reddening in this cluster given its location near the Galactic bulge, there is no evidence of it other than the large scatter in the values of the reddening found by the variety of approaches and samples employed by many authors. In what follows we will adopt the value of $E(B-V) = 0.45$ and discuss what the consequences of adopting lower values have on the cluster distance, the age and the CMD.

\section{The RR Lyrae stars}

\begin{table*}
\begin{center}

\caption{Data of Variable stars in NGC 6171 in the FoV of our images with the exception of V23 whose light curve in the $V$ filter was taken from \citet{Clement1997}.}
\label{variables}

\begin{tabular}{llccccclcc}
\hline
Star ID & Type & $<V>$ (mag) & $<I>$ (mag)   & $A_V$ (mag)  & $A_I$  (mag)   & $P$ [days]   & $\alpha$ (J2000.0)  & $\delta$  (J2000.0)        \\
  
 \hline
V4  & RRc             & 15.493 & 14.667 & 0.462 & 0.343 & 0.282132 & 16:32:25.28 & -13:05:55.4 \\
V5  & RRab \emph{Bl}  & -      & -      & 0.483$^1$ & 0.330$^1$	& 0.702375 & 16:32:47.90 & -13:05:57.3 \\
V6  & RRc \emph{Bl}   & 15.614 & 14.766 & 0.523 & 0.424 & 0.259627 & 16:32:31.32 & -13:04:25.6 \\
V7  & RRab \emph{Bl}  & -      & -      & 1.139$^1$ & 0.746$^1$ & 0.497474 & 16:32:35.00 & -13:04:18.9 \\
V8  & RRab            & 15.578 & 14.523 & 1.094 & 0.668 & 0.559922 & 16:32:32.82 & -13:03:59.9 \\ 
V9  & RRc             & 15.632 & 14.674 & 0.496 & 0.340 & 0.320601 & 16:32:30.18 & -13:03:37.9 \\ 
V10 & RRab \emph{Bl}  & 15.782 & 14.811 & 0.995 & 0.680 & 0.415559 & 16:32:28.08 & -13:03:10.3 \\
V11 & RRab            & 15.673 & 14.543 & 0.680 & 0.444 & 0.592809 & 16:32:32.60 & -13:02:44.9 \\
V12 & RRab \emph{Bl}  & 15.698 & 14.694 & 1.034$^1$ & 0.678$^1$ & 0.472833 & 16:32:35.98 & -13:02:16.6 \\
V13 & RRab            & 15.834 & 14.781 & 1.193 & 0.812 & 0.466797 & 16:32:30.10 & -13:02:06.1 \\
V14 & RRab            & - & - & - &    -                & 0.481620 & 16:32:33.13 & -13:01:55.5 \\
V15 & RRc             & 15.622 & 14.724 & 0.483 & 0.378 & 0.288590 & 16:32:33.20 & -13:01:17.6 \\
V16 & RRab \emph{Bl}  & 15.675 & 14.617 & 0.801$^1$ & 0.485$^1$ & 0.522796 & 16:32:27.33 & -13:01:25.0 \\
V17 & RRab            & 15.593 & 14.538 & 1.058 & 0.681 & 0.561168 & 16:32:25.16 & -13:02:07.2 \\
V18 & RRab            & 15.739 &     -   & 0.811 &    - & 0.561404 & 16:32:37.13 & -12:59:41.9 \\
V19 & RRc \emph{Bl}      & 15.697 & 14.952 & 0.503 & 0.322 & 0.278762 & 16:32:47.79 & -13:00:32.7 \\
V20 & RRab            & 15.708 & 14.610 & 0.789 & 0.537 & 0.578107 & 16:32:34.08 & -13:02:26.8 \\
V21$^2$ & RRc             & 16.519 & 15.660 & 0.566 & 0.407 & 0.258715 & 16:32:37.61 & -13:05:42.5 \\
V23 & RRc             & 15.626 & -      & 0.439 & -     & 0.323344 & 16:32:13.96 & -13:03:01.5 \\
V24 & RRab            & 15.600 & 14.530 & 0.882 & 0.657 & 0.523949 & 16:32:31.98 & -13:03:09.7 \\
V26 & SX Phe          &17.616  & 16.377 & -      & -    &0.055275  &16:32:32.66 & -13:03:00.4  \\
\hline
\end{tabular}
\raggedright
\center{\quad \emph{Bl}: RR Lyrae with Blazhko effect.\\
1. Maximum observed amplitude.\\
2. Not a cluster member}
\end{center}
\end{table*}

\subsection{Fourier decomposition and physical parameters of RR Lyrae stars}
\label{calibrations}

\begin{table*}
\begin{center}

\caption{Fourier coefficients of RRab and RRc stars in NGC 6171 (all in $V$ mag). The numbers in 
parentheses indicate the uncertainty on the last decimal place. Also listed are the reddening values $EBV$ ($E(B-V)$) for each individual RRab star calculated from the method used by \citet{Sturch1966} and the deviation parameter~$Dm$.}     

\label{tab:fourier_coeffs} 
\begin{tabular}{lllllllllll}
\hline
ID     & $A_{0}$    & $A_{1}$   & $A_{2}$   & $A_{3}$   & $A_{4}$  
&$\phi_{21}$
& $\phi_{31}$ & $\phi_{41}$ 
& $EBV$   &$Dm$ \\
     
\hline
\multicolumn{11}{c}{RRab} \\
\hline
V8  & 15.578(1)  & 0.357(1)  & 0.199(1)  & 0.118(1)  & 0.085(1)  & 4.032(10)  & 8.295(14)  & 6.467(20)  & 0.45 & 6.8  \\
V10 & 15.782(1)  & 0.338(13) & 0.156(13) & 0.137(12) & 0.075(11) & 3.530(109) & 7.654(138) & 5.679(211) & 0.37 & 3.7  \\
V11 & 15.673(2)  & 0.229(3)  & 0.120(2)  & 0.064(2)  & 0.037(2)  & 4.010(40)  & 8.426(40)  & 6.750(98)  & 0.50 & 3.4  \\
V12 & 15.698(1)  & 0.429(2)  & 0.169(1)  & 0.064(1)  & 0.030(1)  & 3.922(11)  & 8.066(19)  & 5.748(45)  & 0.43 & 3.3  \\
V13 & 15.834(1)  & 0.415(2)  & 0.194(1)  & 0.135(1)  & 0.081(1)  & 3.816(11)  & 7.954(16)  & 5.834(26)  & 0.46 & 1.5  \\
V16 & 15.675(2)  & 0.292(2)  & 0.153(2)  & 0.073(2)  & 0.028(2)  & 3.896(21)  & 8.143(31)  & 6.264(75)  & 0.47 & 2.9  \\
V17 & 15.593(1)  & 0.340(2)  & 0.188(1)  & 0.115(2)  & 0.075(1)  & 4.024(14)  & 8.396(14)  & 6.554(28)  & 0.43 & 1.4  \\
V18 & 15.739(3)  & 0.282(5)  & 0.143(5)  & 0.084(5)  & 0.046(4)  & 4.162(50)  & 8.355(74)  & 6.494(117) & \multicolumn{1}{c}{-}    & 2.2  \\
V20 & 15.708(1)  & 0.268(2)  & 0.137(1)  & 0.082(1)  & 0.042(1)  & 4.097(20)  & 8.504(25)  & 6.697(40)  & 0.47 & 1.2  \\
V24 & 15.600(16) & 0.287(21) & 0.144(21) & 0.105(20) & 0.071(17) & 3.760(218) & 7.964(306) & 5.850(442) & 0.47 & 2.3  \\
\hline
\multicolumn{11}{c}{RRc} \\
\hline
V4  & 15.493(1) & 0.228(2)  & 0.032(2)  & 0.020(2)  & 0.014(2)  & 4.996(58)   & 3.743(94)  & 2.482(140) &  & \\
V6  & 15.614(1) & 0.250(1)  & 0.049(1)  & 0.013(1)  & 0.023(1)  & 4.429(25)   & 3.817(102) & 1.383(67)  &  & \\
V9  & 15.632(1) & 0.233(1)  & 0.018(1)  & 0.019(1)  & 0.010(1)  & 4.890(48)   & 4.167(50)  & 2.554(90)  &  & \\
V15 & 15.622(1) & 0.240(2)  & 0.032(2)  & 0.019(2)  & 0.013(2)  & 4.838(49)   & 3.596(82)  & 2.114(125) &  & \\
V19 & 15.697(1) & 0.215(1)  & 0.042(1)  & 0.013(1)  & 0.012(1)  & 4.137(31)   & 2.255(97)  & 2.142(109) &  & \\
V21 & 16.519(8) & 0.230(11) & 0.047(11) & 0.012(11) & 0.012(11) & 4.638(248)  & 3.129(855) & 1.099(920) &  & \\
V23 & 15.626(1) & 0.207(2)  & 0.019(2)  & 0.013(2)  & 0.009(2)  & 4.783(101)  & 3.947(146) & 2.505(198) &  & \\
\hline
\label{parfou}
\end{tabular}
\end{center}
\end{table*}

\begin{table*}
\begin{center}

\caption{Physical parameters of the RRab and RRc stars in NGC 6171. The numbers in parentheses 
indicate the uncertainty on the last decimal place.} 
\label{ParFis}

 \begin{tabular}{clllllll}
\multicolumn{8}{c}{RRab} \\
\hline 
ID&[Fe/H]$_{\rm ZW}$ & [Fe/H]$_{\rm UVES}$ &$M_V$ & log~$T_{\rm eff}$  &log$(L/{
L_{\odot}})$ &$M/{ M_{\odot}}$&$R/{ R_{\odot}}$\\
\hline

\hline
V8$^1$ &-1.40(13)  &-1.30(13)  &0.566(1)  &3.811(7)  &1.673(1)  &0.69(6)  &5.50(1)  \\
V11 &-1.40(38)  &-1.30(38)  &0.623(4)  &3.804(13) &1.651(2)  &0.65(10)  &5.54(1) \\
V13 &-1.37(15)  &-1.26(15)  &0.661(2)  &3.821(7)  &1.635(1)  &0.72(6)  &5.03(1)  \\
V17 &-1.31(13)  &-1.20(12)  &0.581(3)  &3.812(8)  &1.668(1)  &0.67(6)  &5.44(1)  \\
V18 &-1.35(70)  &-1.24(68)  &0.623(7)  &3.810(15) &1.651(3)  &0.66(12) &5.40(2) \\
V20 &-1.27(23)  &-1.16(22)  &0.614(2)  &3.809(8)  &1.655(1)  &0.64(6)  &5.44(1)  \\
V24 &-1.58(288) &-1.50(325) &0.690(29) &3.810(50) &1.624(12) &0.67(40) &5.21(7) \\
 
\hline
Weighted mean& -1.33(1)&-1.22(1) &0.596(1) &3.813(4) & 1.661(1)&0.68(3)&5.34(1)\\
$\sigma$&$\pm$ 0.12 &$\pm$ 0.12 & $\pm$ 0.051 &$\pm$ 0.002 &$\pm$ 0.006&$\pm$ 0.01 &$\pm$ 0.07\\
\hline

\\

\multicolumn{8}{c}{RRc} \\
\hline
ID&[Fe/H]$_{\rm ZW}$ & [Fe/H]$_{\rm UVES}$ &$M_V$ & log~$T_{\rm eff}$  &log$(L/{
L_{\odot}})$ &$M/{ M_{\odot}}$&$R/{R_{\odot}}$\\
\hline

\hline
V4  &-0.81(17) &-0.75(10)  &0.577(9)  &3.875(1) &1.669(4) &0.57(1) &4.07(1) \\
V9  &-1.07(10) &-0.96(8)   &0.562(5)  &3.870(1) &1.675(2) &0.51(1) &4.21(1) \\
V15 &-1.02(15) &-0.91(11)  &0.582(9)  &3.873(1) &1.667(4) &0.56(1) &4.11(2) \\
V21$^2$ & -0.89 (1.43)& -0.81(92)  &0.624(50) & 3.877(5)& 1.650(20)&0.60(5)& 3.95(9)\\
V23 &-1.26(29) &-1.14(26)  &0.569(10) &3.868(1) &1.672(4) &0.51(1) &4.23(2) \\
\hline
Weighted mean&-1.02(7)& -0.90(5)&0.569(3) &3.871(1) &1.672(1) &0.60(1) & 4.13(1)\\
$\sigma$&$\pm$ 0.19 &$\pm$ 0.19 &$\pm$ 0.010 &$\pm$ 0.002 &$\pm$ 0.002 &$\pm$ 0.02 &$\pm$ 0.04\\
\hline

\label{parfis}
\end{tabular}
\center{1. Not included in the average of [Fe/H]. \\
2. Not a cluster member, hence not included in the calculation of mean physical parameters.}
\end{center}
\end{table*}

Since RR Lyrae stars are the most relevant stars in globular clusters and are good proxies of the stellar evolution on the HB, it is useful to determine their physical parameters such as [Fe/H], log($L/L_{\odot}$), log $T_{\rm eff}$, mass and radius. This can be achieved by means of the Fourier decomposition of their light curves in $V$ into harmonics and by using semi-empirical calibrations that can correlate the Fourier parameters with the physical quantities. The $V$ and $I$ light curves of the RR Lyrae stars are shown in Fig. \ref{mosaico_6171}.

A given light curve of a RR Lyrae star can be represented by the following equation: 
\begin{equation}
\label{eq.Foufit}
m(t) = A_0 + \sum_{k=1}^{N}{A_k \cos\ ({2\pi \over P}~k~(t-E) + \phi_k) },
\end{equation}

\noindent
where $m(t)$ is the magnitude at time $t$, $P$ is the period of pulsation, and $E$ is the epoch. In order to calculate the Fourier parameters, a least-squares approach is used to estimate the best fit for the amplitudes $A_k$ and phases $\phi_k$ of the sinusoidal components.  The phases and amplitudes of the harmonics in Eq.~\ref{eq.Foufit}, i.e. the Fourier parameters, are defined as $\phi_{ij} = j\phi_{i} - i\phi_{j}$, and $R_{ij} = A_{i}/A_{j}$.

Well tested calibrations and zero points have been used systematically by our group over the last few years \citep{Are2017}. Since other calibrations have been used by other authors, below we list the specific equations used in this work. For metallicity and absolute magnitude of the RRab stars the calibrations of \cite{Jurcsik1996} and \cite{Kovacs2001} were employed, respectively:\\

\begin{equation} 
{\rm [Fe/H]}_{J} = -5.038 ~-~ 5.394~P ~+~ 1.345~\phi^{(s)}_{31},
\label{eq:JK96}
\end{equation} 

\begin{equation} 
M_V = ~-1.876~\log~P ~-1.158~A_1 ~+0.821~A_3 + K.
\label{eq:ḰW01}
\end{equation} 

\noindent  
Note that the calibration for the metallicity is given in the Jurcsik-Kov\'acs scale, but can be transformed to the Zinn-West scale \citep{Zinn1984} with the equation: 
[Fe/H]$_{J}$ = 1.431[Fe/H]$_{ZW}$ + 0.88
\citep{Jurcsik1995}. \\

For the RRc stars we employed the calibrations given by  \cite{Morgan2007} and \cite{Kovacs1998}, respectively:

\noindent

$${\rm [Fe/H]}_{ZW} = 52.466~P^2 ~-~ 30.075~P ~+~ 0.131~\phi^{(c)~2}_{31}$$
\begin{equation}
~~~~~~~	~-~ 0.982 ~ \phi^{(c)}_{31} ~-~ 4.198~\phi^{(c)}_{31}~P ~+~ 2.424,
\label{eq:Morgan07}
\end{equation}

\begin{equation}
M_V = 1.061 ~-~ 0.961~P ~-~ 0.044~\phi^{(s)}_{21} ~-~ 4.447~A_4.
\label{eq:K98}	
\end{equation}

\noindent 

The coefficients can be transformed from cosine series phases into sine series via the relation: ~~ $\phi^{(s)}_{jk} = \phi^{(c)}_{jk} - (j - k) {\pi \over 2}$, when necessary. 

In Table \ref{parfou} the Fourier parameters of the RR Lyrae identified in NGC 6171 are listed and in Table \ref{parfis} the physical parameters derived from the Fourier decomposition are reported. It is worth recalling that the determination of the iron abundance [Fe/H] of the RRab stars is only valid if they comply with the {\it Deviation Parameter} or $Dm$ \citep{Jurcsik1996} not exceeding the value of 3.0.
The $Dm$ value for each RRab star is listed in Table \ref{parfou} in column 11. Since a few stars have $Dm$ only marginally larger, we relaxed the criterion in order to include these stars and have a larger sample. For V8, with $Dm =6.8$, its Fourier parameters are reported but were not taken into account when calculating the average of [Fe/H]. Also not included were the variables V5, V6, V7, V12, V16 and V19 since they show Blazhko-like amplitude modulations, and V14 that has an incomplete light curve. Despite V21 being a non-member, we calculated its physical parameters. 

The iron abundance in the scale of \cite{Zinn1984} can be converted into the scale of \cite{Carretta2009} via the equation [Fe/H]$_{\rm UVES} = -0.413 + 0.130$[Fe/H]$_{\rm ZW} -0.356$[Fe/H]$^2_{\rm ZW}$, and are also listed in Table \ref{ParFis}.

With the period, luminosity and temperature known for each
RR Lyrae star, its mass and radius can be estimated from the equations:
log~M/M$_{\odot}$
 = 16.907 -- 1.47 log~$P_F$ + 1.24 log (L/L$_{\odot}$) -- 5.12 log~$T_{\rm eff}$ \citep{vAlbada1971} and $L = 4 \pi R^2 \sigma T^4$, respectively.

\subsection{Bailey diagram and Oosterhoff type}
\label{secBailey}

\begin{figure}
\includegraphics[width=6cm, height=10cm]{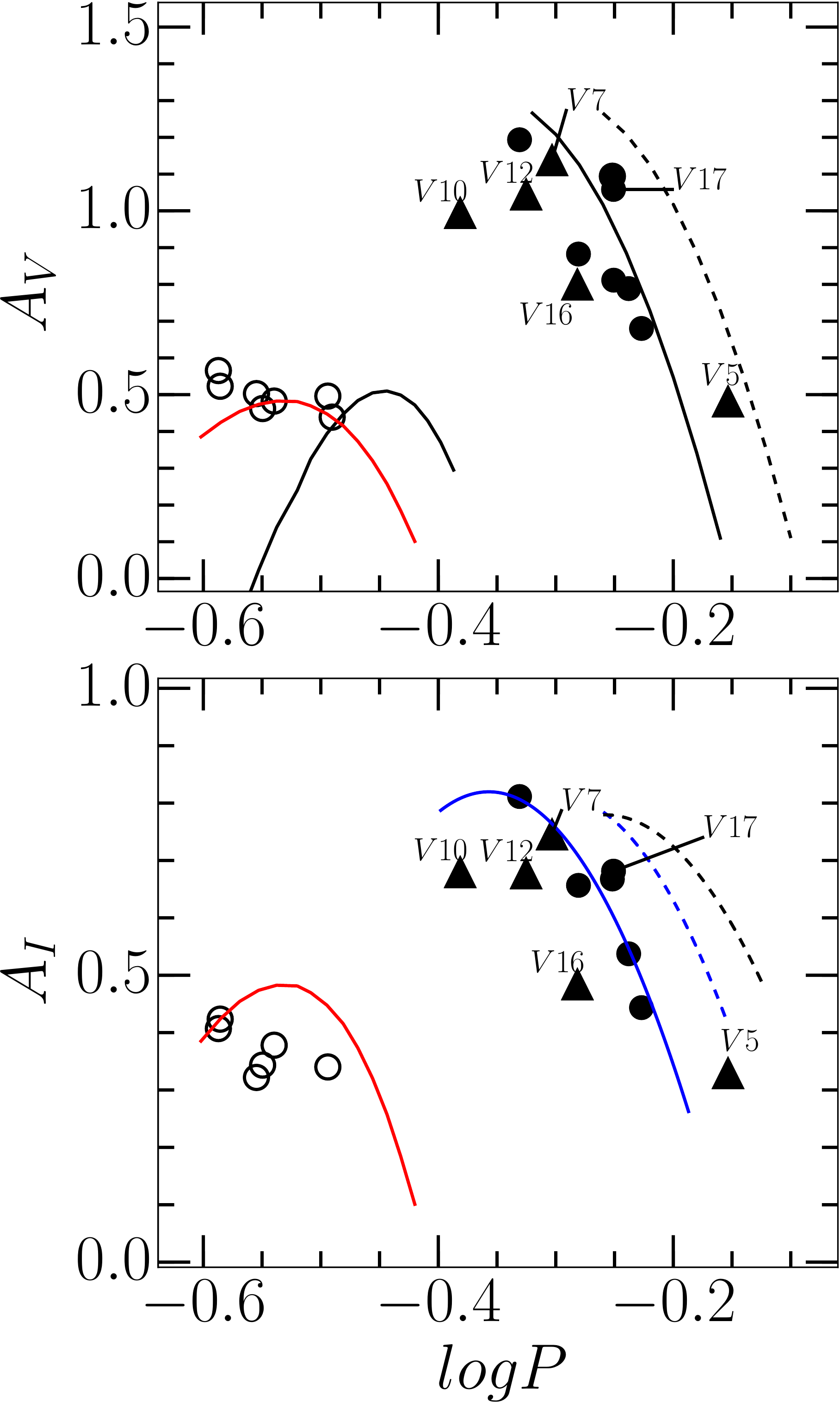}
\caption{Bailey diagram for NGC 6171. Filled and open symbols represent RRab and RRc stars, respectively. Triangles correspond to stars with Blazhko modulations. The continuous and segmented lines in the top panel are the loci for unevolved and evolved stars in M3 according to \cite{Cacciari2005}. The black parabola was obtained by \cite{Kunder2013a} for RRc stars in 14 OoII clusters. The red parabolas were calculated by \cite{Arellano2015} from a sample of RRc stars in five OoI clusters and avoiding Blazhko variables. In the bottom panel the black segmented locus was found by \cite{Arellano2011} and \cite{Arellano2013} for the OoII clusters NGC 5024 and NGC 6333 respectively. The blue solid and segmented loci for unevolved and evolved stars respectively are from \citet{Kunder2013b}. For a more detailed discussion, see $\S$ \ref{secBailey}.}
\label{figBailey} 
\end{figure}

The period-amplitude plane for RR Lyrae stars, also known as the Bailey diagram, is shown in Fig. \ref{figBailey} for the \emph{VI} band passes. Periods and amplitudes are listed in Table 2. In most cases, we took the amplitudes corresponding to the best fit provided by the Fourier decomposition of the light curves unless there were glaring gaps such as in the case of V14 or where amplitude and phase modulations were significant, like in V5 and V7. In these cases the star was ignored or, in other cases, the maximum amplitude was measured and the star was plotted with a triangle in Fig. \ref{figBailey}. With the exception of these Blazhko variables, the distribution of RRab stars in both \emph{VI} band passes shows a mild scatter around the OoI locus found by \citet{Cacciari2005} for M3 and by \cite{Kunder2013a} for NGC 2808. This confirms the OoI type of NGC 6171. While averaging the periods of the RRab stars, we found $<P_{ab}>$ = 0.53d which also defines NGC 6171 as of the OoI type. None of the RRab stars falls near the evolved sequences, i.e. the segmented loci, thus none of them 
seems to be clearly ahead in its evolution towards the asymptotic giant branch. V17 was found to have a significant positive secular period change \citep{are2018}, expected in stars advanced in their evolution. However, while in the $V$-band the scatter is larger and it may not preclude V17 from being inconsistent with the evolved stars locus, in the $I$-band all stable RRab stars are rather consistent with the non-evolved sequence.

\begin{figure} 
\includegraphics[width=8.0cm,height=4.3cm]{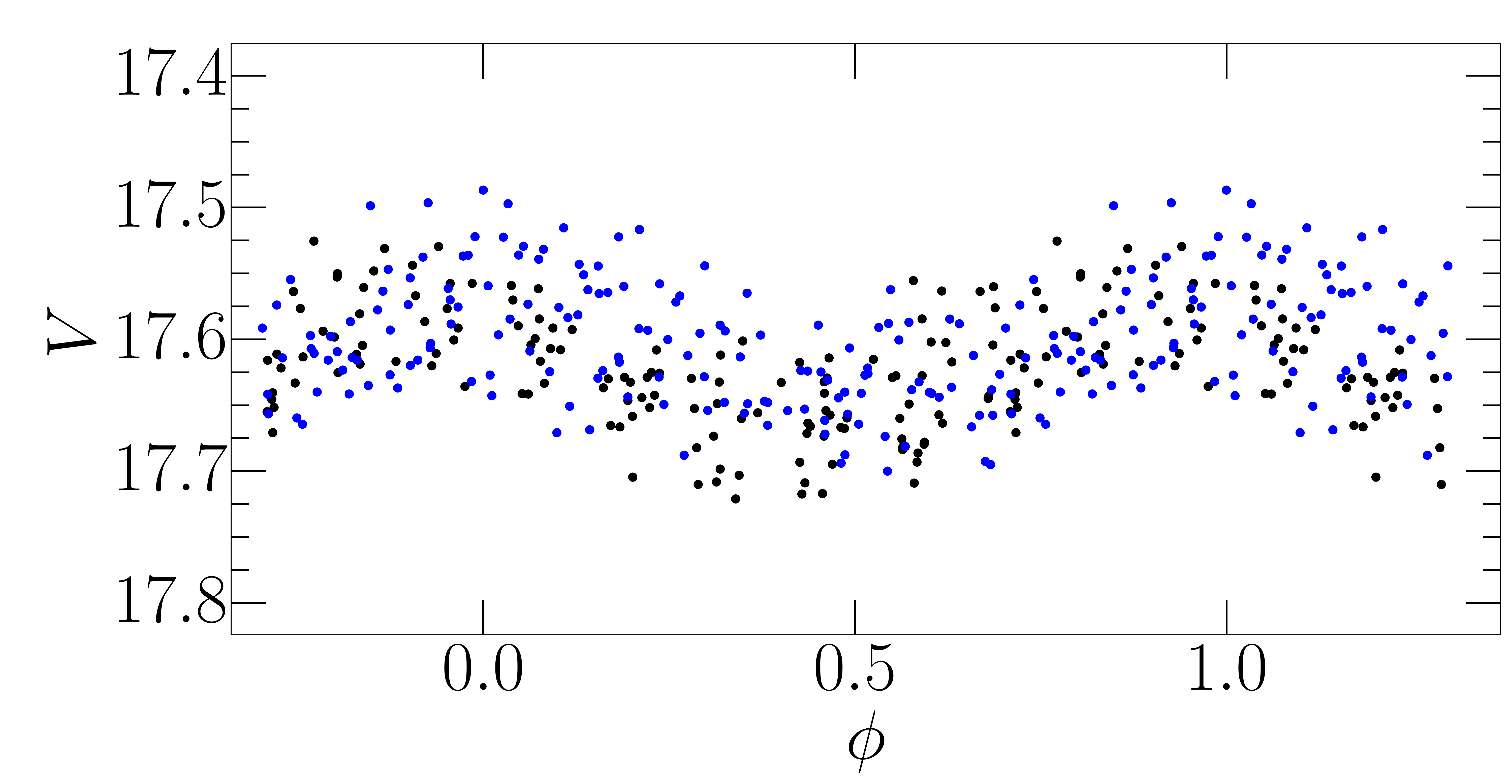}
\caption{Light curve of V26. The black and blue dots correspond to the data from Hanle and San Pedro M\'artir respectively.}
    \label{V26}
\end{figure}

\section{The [Fe/H] value of NGC 6171}
\label{metalicity}

NGC 6171 is a moderately metal rich cluster of the OoI type. In Table \ref{metallicity} we list various metallicity values found in the literature and the methods used to determine them. We include the results from Fourier decomposition found in this work for comparison. We can see that many of the results come from the spectroscopic metallicity index $\Delta S$ defined by \citet{Preston1959}. These values agree well within the uncertainty with the value derived for the RRc stars in the present work. The result for the RRab stars, [Fe/H]= $-1.33 \pm 0.12$, although in a reasonable agreement with the other estimates, is the lowest value. Nevertheless, it is in excellent agreement with the 
trend of the $Mv$-[Fe/H]$_{ZW}$ relation (see Fig. \ref{MVFEH:6171}) found by \cite{Are2017} for the RRab stars from a sample of 24 globular clusters.

\begin{table}[h!]
\begin{center}

\caption{ Historical values of [Fe/H] for NGC 6171.}
\label{metallicity}
\begin{tabular}{lc}
\hline
Method & [Fe/H] \\
\hline
 Spectroscopic metallicity index $\Delta S$       & -0.82$^1$   \\
 High resolution spectroscopy                     & -1.00$^2$    \\
 Spectroscopic metallicity index $\Delta S$       & -0.83, -0.94$^3$    \\
 Spectroscopic metallicity index $\Delta S$       & -0.84$^4$   \\
 Mean from previous $\Delta S$ determinations     & -0.90$^5$   \\
 Webbink's empirical relation  & -1.09$^6$   \\
 Revised $\Delta S$ method     & -0.99$^7$	\\
 Spectroscopic analysis                           & -0.93$^8$   \\ 
 RRab Fourier decomposition                       & -1.33$^9$	\\
 RRc Fourier decomposition                        & -1.02$^9$	\\

\hline
\end{tabular}
\raggedright
\center{\quad 1: \citet{Zinn80}, 2: \citet{Pilachowski1981}, 3: \citet{Smith1982}, 4: \citet{Smith1983}, 5: \citet{DaCosta1984}, 6: \citet{Cudworth1992}, 7: \citet{Clementini2005}, 8: \citet{OConnell2011}, 9: This work.}
\end{center}
\end{table}

\begin{figure*}
\begin{center}
\includegraphics[width=15cm, height=15cm]{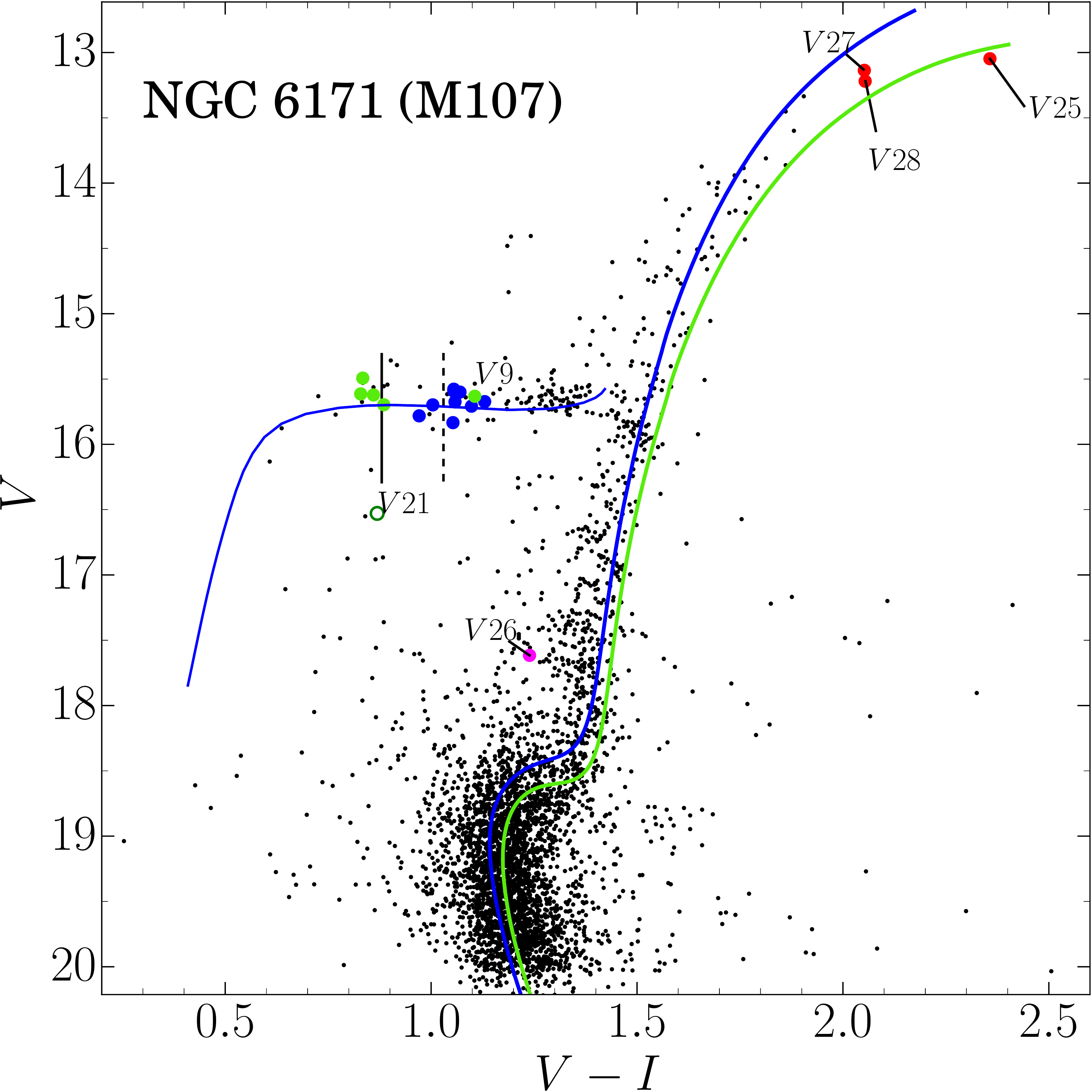}
\caption{CMD of NGC 6171. Blue circles correspond to RRab stars, green circles to RRc stars and the magenta circle to the SX Phe. The open green circle corresponds to a RRc star (V21) that is not a member of the cluster. 	V27 and V28 are variables newly found in the present work. The two isochrones are from \cite{Vandenberg2014}, with  Y = 0.25 and [$\alpha$/H] = 0.4, corresponding to an age of 11 Gyr and to a [Fe/H]$_{RRab}$ = -1.33 (blue) and to a [Fe/H]$_{RRc}$ = -1.02 (green).  The ZAHB for [Fe/H] = -1.31 and isochrones have been reddened by $E(B-V)=0.45$ and shifted to the distance from the RRab and RRc stars, i.e. 5.3 kpc. The continuous and dashed vertical lines represent the red edge of the first overtone instability strip
reddened by $E(B-V)$ 0.33 and 0.45 respectively. See $\S$ \ref{sec:CMD} for a discussion.}
\label{CMD_6171}
\end{center}
\end{figure*}

As a further test, we applied eqs. \ref{eq:JK96} and \ref{eq:Morgan07} to the
Fourier parameters published by \cite{Clement1997} for the stable members of NGC 6171 and found  [Fe/H]$_{ZW}$ -1.31 $\pm$ 0.02 and -1.03 $\pm$ 0.04 for the RRab and RRc stars respectively, which are very similar to the results from the decomposition of our own light curves in the present work.

We note that the value from the RRc stars is similar to the spectroscopic results.

\section{The Distance to NGC 6171} 
\label{Distance}

The estimation of the distance to globular clusters relies heavily upon an accurate determination of the reddening. For NGC 6171 the value of the distance and reddening that was adopted by \citet{Harris1996} are 6.4 kpc and $E(B-V) = 0.33$ respectively. However, as discussed in $\S$ \ref{reddening}, the value of $E(B-V)$ from a large number of independent estimates has been found to have a large scatter. From the average reddening calculated from the minimum colour in RRab stars we have adopted $E(B-V) = 0.45$.

To estimate the distance to NGC 6171 we made use of four independent methods, all of which yield a consistent value within their respective uncertainties. We describe these methods in the following sections.  

\subsection{From the RR Lyrae stars}

For the RR Lyrae stars the values of absolute magnitude $M_V$ of each star were employed to calculate individual distances. Since two independent calibrations for RRab and RRc stars for $M_{V}$ were used, then two equally independent values of the distance to NGC 6171 were found; 5.3 $\pm$ 0.3 kpc and 5.3 $\pm$ 0.2 kpc for the RRab and RRc stars respectively. We stress at this point that the individual values of $E(B-V)$ listed in Table \ref{tab:fourier_coeffs} for the RRab stars were employed in the calculations. For the RRc stars, given that we do not know individual reddenings, we adopted the mean from the RRab stars, $E(B-V) = 0.45$.

An alternative approach is via the P-L relation for RR Lyrae stars in the $I$ filter \citep{Catelan2004} $M_I = 0.471-1.132~ {\rm log}~P +0.205~ {\rm log}~Z$, with ${\rm log}~Z = [M/H]-1.765$; $[M/H] = \rm{[Fe/H]} - \rm {log} (0.638~f + 0.362)$ and log~f = [$\alpha$/Fe] \citep{Sal93}. Thus, adopting a value for [Fe/H] and a reddening one can obtain the corresponding average distance. Applying this relation to the RRab and RRc stars, each with their [Fe/H]$_{\rm ZW}$ value as given in Table \ref{ParFis} and adopting [$\alpha$/Fe]=+0.3, we found the average distance $<d>$ = 5.6 $\pm$ 0.2 kpc  and $<d>$ = 5.4 $\pm$ 0.3 kpc respectively, that are consistent with the distance values obtained from the Fourier approach described above. 
RR Lyrae stars with either Blazhko amplitude modulations (V5, V6, V7, V12, V16, V19), incomplete $I$ light curve (V14, V18), no  $I$ data available (V23) or not a member of the cluster (V21), were ignored. 

\subsection{From the SX Phe star}

There is only one known SX Phe star in NGC 6171, V26. The period and the intensity weighted $<V>$ and $<I>$ as well as other relevant data for this star are listed in Table \ref{variables}.
In order to calculate the distance to NGC 6171 from this SX Phe, we adopted the P-L relation for SX Phe stars from \citet{CohenSara2012} for the fundamental mode $M_v =(-1.64 \pm 0.11) - (3.39 \pm 0.09)~log~P_f$. As we did with the RR Lyrae stars, we also adopted the $E(B-V)$ = 0.45. 

The resulting distance to V26 is $d$ = 5.2 $\pm$ 0.4 kpc, in agreement with the distance obtained above from the RR Lyrae stars. This confirms that V26 is a fundamental pulsator and a member of the cluster. The position of V26 in the CMD diagram is shown in Fig. \ref{CMD_6171}. Since there is only one SX Phe, the quoted uncertainty corresponds to the uncertainty associated to the P-L relation.

It is worth noting that the light curve of the SX Phe is very disperse, due to the fact that the star is faint and it is located near the central region of the cluster. We found no traces of a secondary pulsating mode.

The weighted mean of the distance rendered by the above four approaches is 5.4 $\pm$ 0.1 kpc. In Table \ref{distance} we compare this result with some values from the literature.

\begin{table}[h!]
\begin{center}

\caption{Other values of the distance to NGC 6171.}
\label{distance}
\begin{tabular}{lc}
\hline
Method & Distance [kpc]      \\
\hline
 Assuming $Mv$ = +0.5 for the HB.   & 7.2$^1$       \\
Absolute and apparent magnitude of the HB   & 6.2 $\pm$ 0.9$^2$ \\
From bolometrically corrected magnitudes         & 5.6$^3$       \\
Fourier decomposition; P-L RRL, SX Phe  & 5.4 $\pm$ 0.1$^4$   \\

\hline
\end{tabular}
\raggedright
\center{\quad 1: \citet{Peterson1975},  2: \citet{Cudworth1992}, 3: \citet{Shetrone2009}, 4: This work.}
\end{center}
\end{table}

\subsection{The tip of the red giant branch as a distance indicator}

The luminosity of the true tip of the RGB (TRGB) can in principle be used to estimate the distance to globular clusters. It has been argued on theoretical grounds that the neutrino magnetic dipole moment enhances the plasma decay process, postpones helium ignition in low-mass stars, and extends the red giant branch, hence the brightest stars in a given cluster may lay below the true TRGB. \cite{Viaux2013} found that the difference for the case of M5 is between 0.04 and 0.16 mag. By a comparison of their non-canonical models with the observed TRGB in 25 clusters, \cite{Arceo2015} concluded that in average the theoretical TRGB is some 0.26 bolometric magnitudes brighter than the observed one with a standard deviation of 0.24 (see their Table 5).

For the case of NGC 6171, we note from the CMD in Fig. \ref{CMD_6171}, that the three brightest stars are about 0.98 mag fainter than the theoretical TRGB as calibrated by \cite{SalCas1997};

\begin{equation}
\label{TRGB}
M_{bol}^{tip} = -3.949\, -0.178\, [M/H] + 0.008\, [M/H]^2,
\end{equation}

However, it is likely that our photometry for this bright stars is still affected by systematic errors that could not be fully removed by the post calibration described in $\S$ \ref{calibration}, and hence, the TRGB bends towards lower luminosities, as suggested by the comparison with the isochrones (see $\S$ \ref{sec:CMD} below). Therefore, this approach does not lead to a reliable independent estimate of the cluster distance.

\section{The Colour-Magnitude Diagram and the age of NGC 6171} 
\label{sec:CMD}

The CMD of NGC 6171 in Fig. \ref{CMD_6171} was built using the magnitude weighted means of $V$ and $V - I$ of the 3827 measured stars in the FoV of our images. For all variables, the intensity-weighted means $<V>$ and $<V>-<I>$ were used. These were calculated from the Fourier fits of light curves in Fig. \ref{mosaico_6171} and are listed in Table \ref{variables}.

In order to confirm that the CMD is consistent with the basic parameters found in this work, i.e. reddening, metallicity and distance we overlaid theoretical isochrones and a Zero Age Horizontal Branch (ZAHB) taken from the Victoria-Regina stellar models of \cite{Vandenberg2014} kindly made available by the authors. For the age of the cluster we refer to differential age estimation by \cite{DeAngeli2005} from which an age of 10.98 Gyrs can be inferred for NGC 6171. We adopted the isochrones corresponding to 11.00 Gyrs, and for [Fe/H] = -1.33 (blue) and -1.02 (green) as suggested from the RRab and RRc stars respectively. The ZAHB for [Fe/H] = -1.31 is also shown. These models were reddened by $E(B-V)=0.45$ and shifted to the distance of 5.3 kpc and produce a satisfactory match with the observed CMD.

We must stress at this point that, if the alternate values of $E(B-V)=0.33$ and distance 6.4 kpc are used, both the isochrones and the ZAHB would be way off to the red of the observed stellar distributions.

The distribution of RRab and RRc in the instability strip on the HB should be noted. Except for V9, the two modes seem to be neatly separated, as it occurs in all OoII type clusters but only in some OoI clusters \citep{are2018}. The split border
between modes was calibrated by \cite{Arellano2016} at $(V-I)_0 = 0.45-0.46$ and was interpreted as the red edge of the first overtone (RFO) instability strip. Hence no RRc stars should be found to the red side of this border. In Fig \ref{CMD_6171} this border, reddened by 0.33 and 0.46 is shown as a continuous and segmented vertical lines respectively.  In order to place RFO in the observed separation of modes, a reddening of between 0.35-0.40 would be 
required. We believe that all the disagreements in the value of reddening, and those commented in $\S$ \ref{reddening}, may suggest the presence, to some extent, of differential reddening, which given the Galactic position of the cluster should not come as a surprise. 

Let us turn our discussion towards the RRc variable V9, located beyond the red edge of the first overtone of the instability strip. Its position in the CMD is firm given the high photometric quality of the $V$ and $I$ light curves. One possible explanation can be based upon higher reddening than average value. If we consider that the star should be situated somewhere between two RFO's in Fig. 5, a reddening value higher by 0.08-0.12 than the average reddening would be required to reconcile its location in the CMD. This would imply that the star in the present position is dimmer by 0.25-0.35 magnitude. While advocating the possibility of the presence  of differential reddening in NGC 6171, we should refrain from further  speculative discussion on this star.

In spite of the above reddening demeanors we consider NGC 6171 as one more example of a very red HB morphology ((B-R)/(B+V+R) = --0.74) OoI cluster with the two pulsation modes well split.

\section{Discussion}
In this paper we present the results of the first ever determination of physical parameters of RR Lyrae stars in NGC 6171 by means of Fourier decomposition of their light curves. By performing high-precision photometry we were able to extract the light curves of 3827 stars in our reference image, 20 of which are RR Lyrae, one SX Phe and three Lb variables. Of the three Lb variables, the variability of two is newly announced in this work (V27 and V28).

\begin{figure}
\includegraphics[width=8cm, height=4cm]{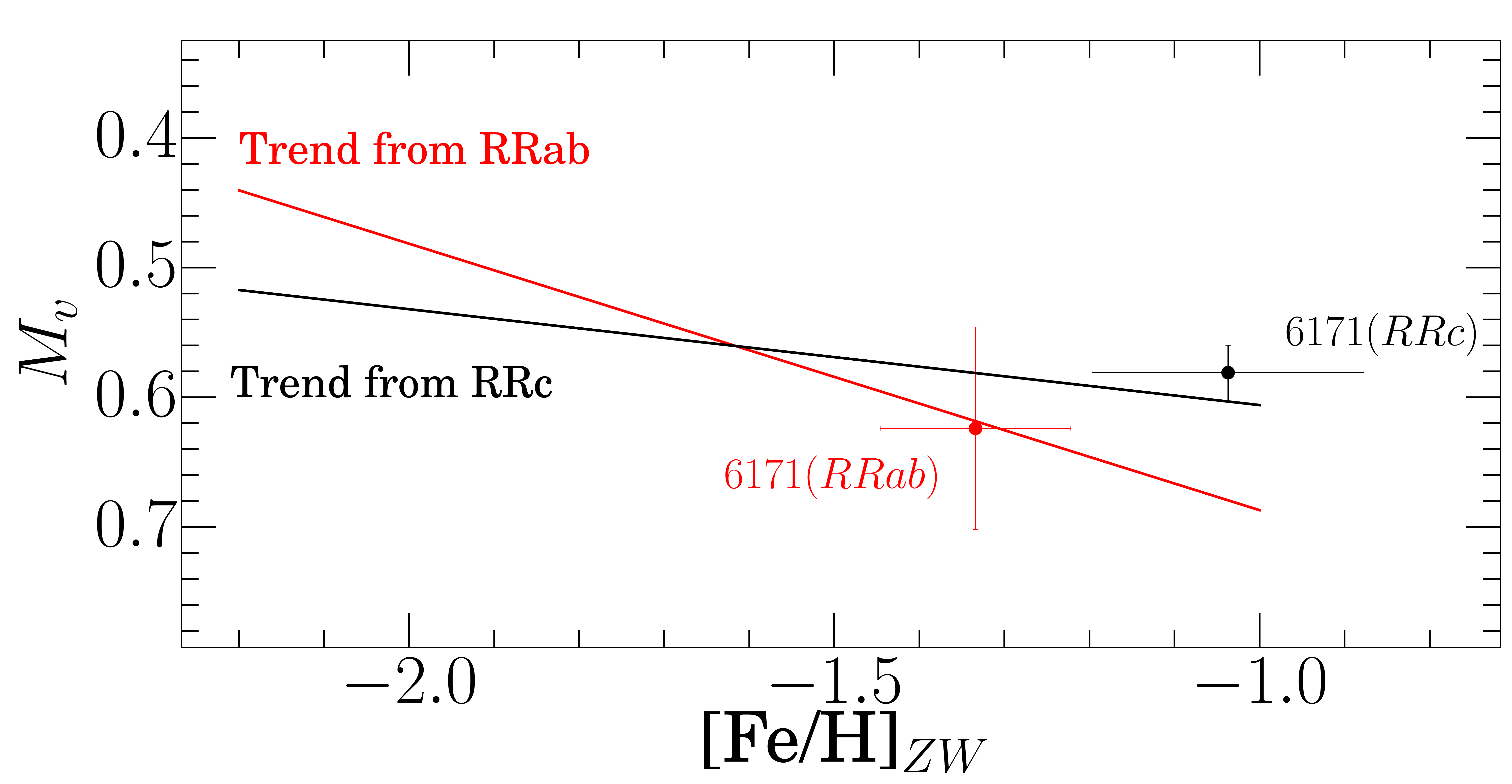}
\caption{Position of NGC 6171 on the [Fe/H]-$M_V$ plane according to the Fourier decomposition results of RRab and RRc  stars. The lines are the trends found by \cite{Are2017} for a family of 24 globular clusters.}
\label{MVFEH:6171}
\end{figure}

An estimation of the reddening via the minimum colour of RRab stars suggests a mean value of $E(B-V)=0.45$. However, historical determinations of $E(B-V)$ display a large scatter between 0.25 and 0.48. This fact may seem to suggest that some differential reddening may be present. Taking individual reddening values for the RRab stars and an average of 0.45 for the RRc stars, and considering the results from the P-L relations for the RR Lyrae and SX Phe stars, we found an overall average distance of 5.4 $\pm$ 0.1 kpc. For the RRab and RRc the metallicities are [Fe/H]$_{ZW}$ $-1.33 \pm 0.12$ and -1.02 $\pm$ 0.19 respectively. It is worth noticing that the result from the RRc stars are closer to the spectroscopic results which are indisputably the most accurate and are not affected by the reddening.

The above distance and iron values tend to be to the lower end of the range of previous estimates, as shown in Tables 6 and 7. Particularly [Fe/H]$_{ZW}$ of $-1.33$ from the RRab seems a bit too low. We should highlight however that the isochrone for this metallicity fits the observations very well. If the isochrone for [Fe/H]$_{ZW} = -1.02$ from the RRc stars is shifted upwards to match the data, it would imply an even shorter distance of 5.0 kpc.

Let us refer to the $Mv$-[Fe/H]$_{ZW}$ relation for RR Lyrae. From the homogeneous Fourier decomposition of RR Lyrae stars in a sample of 24 clusters of both OoI and OoII types, \cite{Are2017} found that the slope of the relationship is different when calculated from RRab and from RRc stars. In Fig. \ref{MVFEH:6171} the two trends for the RRab and RRc stars are shown and the corresponding position of NGC 6171 given the values in Table \ref{ParFis}. This shows that despite the difference in the metallicity from the two stellar groups, the present results are consistent with the homogeneous Fourier analysis in a larger sample of globular clusters.

\acknowledgements
We acknowledge the financial support from DGAPA-UNAM via grants 
IN106615-17, IN105115 and from CONACyT (M\'exico). DD is grateful to CONACyT, M\'exico for the PhD scholarship.
We have made an extensive use of the SIMBAD and ADS services, for which we are thankful.

\appendix

\section{Comments on individual variables}
\label{individual}

\subsection{V5, V6, V7, V12, V16, V19}
All these stars show clear signs of amplitude and/or phase modulations.

\begin{table}[!th]
\begin{center}
\caption{Irregular variables giants Lb in  NGC 6171.}
\label{SRs}
\begin{tabular}{ccccc}
\hline
ID &  $<V>$  & $<I>$  & $\alpha$ & $\delta$ \\
 &(mag) & (mag) &  (J2000.0)& (J2000.0)\\
\hline
V25    & 13.048 &10.691    &16:32:31.41  & -13:02:59.1  \\
V27    & 13.137 &11.084    &16:32:24.00   & -13:02:10.4  \\
V28    & 13.220 &11.166    &16:32:39.00   & -13:02:50.5  \\
\hline
\end{tabular}
\raggedright
\end{center}
\end{table}

\subsection{V10}
Although the general distribution of RR Lyrae stars  in the period - amplitude plane follows that of an OoI cluster, V10 shows peculiar position despite having a well-defined light curve as noted in $\S$ \ref{secBailey}. \cite{Clement1997} find a mild irregularity in the light curve of V10 claiming that its shape is not consistent from cycle to cycle. They attribute this behaviour to a probable Blazhko effect, which would explain the off-trend position of this star and also would make it consistent with the position of variables V12 and V16 in the Bailey diagram.  A close inspection of the light curve  of V10 published by \citet{are2018}, indeed seems to indicate a mild amplitude modulation. \citet{are2018} report that this star is going through a secular period decrease.

\subsection{V19}

\cite{Kovacs1986} noted this star as a double-mode pulsator but the first overtone to fundamental mode period ratio P1/P0 was not found to be the canonical for radial pulsations. Given that the light curve presented in this work is very well defined over the time span of two years, we performed a frequency analysis and we did not detect a significant secondary frequency.

\subsection{V21} 
This RRc star is in fact farther than the cluster as it falls well below the HB in the CMD of Fig. \ref{CMD_6171}. Its distance, from the calibration described in  $\S$ \ref{calibrations} is 7.8 kpc. The non-member status of this star was first noted by \cite{Dickens1970} and later confirmed by \cite{Cudworth1992}. 

\subsection{V25, V27 and V28}
\begin{figure*}
\includegraphics[width=17cm, height=9cm]{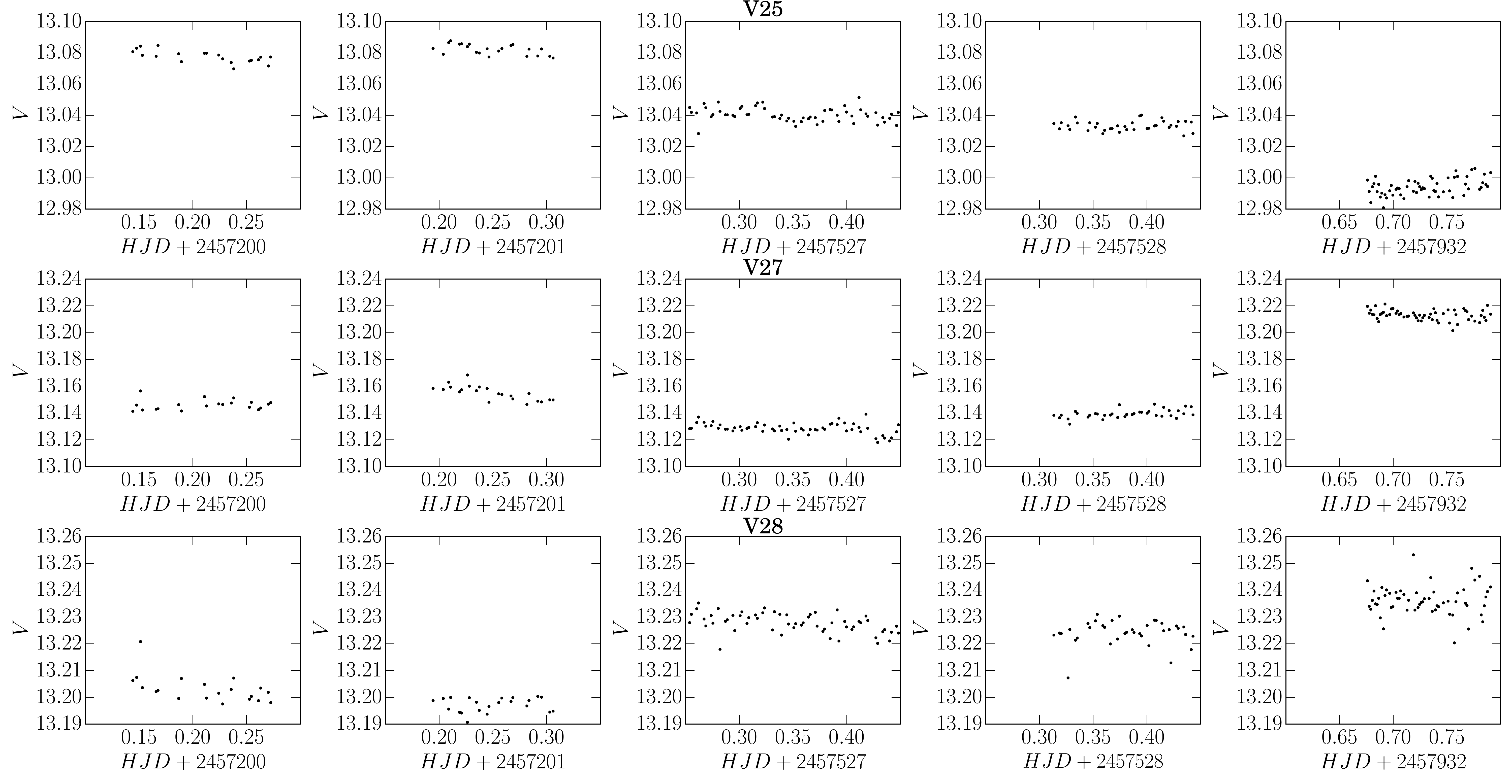}
\caption{Long term variability in three Lb type stars. V27 and V28 which are newly announced in the present paper.}
\label{fig:SRs}
\end{figure*}

{\bf V25}. The variability of this star was reported by \cite{Lloyd1973} and \cite{Lloyd1977}. They  acknowledge that their classification as variable depends only on one $V$ and one $I$ plates. The star was classified as Lb by \cite{Clement2001} in their CVSGC. On the basis of our present data we confirm the irregular long term variable nature of this star and this is shown in Fig. \ref{fig:SRs}. \\

\noindent
{\bf V27, V28}. We report here the variability of these two stars. Like V25, they present long-term irregular variations, hence we classify them as Lb variables. Their light curves are displayed in Fig. \ref{fig:SRs}. The mean magnitudes and equatorial coordinates of V25, V27 and V28 are given in Table \ref{SRs}. Their position near the TRGB can be seen in the CMD of Fig. \ref{CMD_6171}.  

\begin{figure*}[!th]
\begin{center}
\includegraphics[width=11.50cm, height=11.50cm]{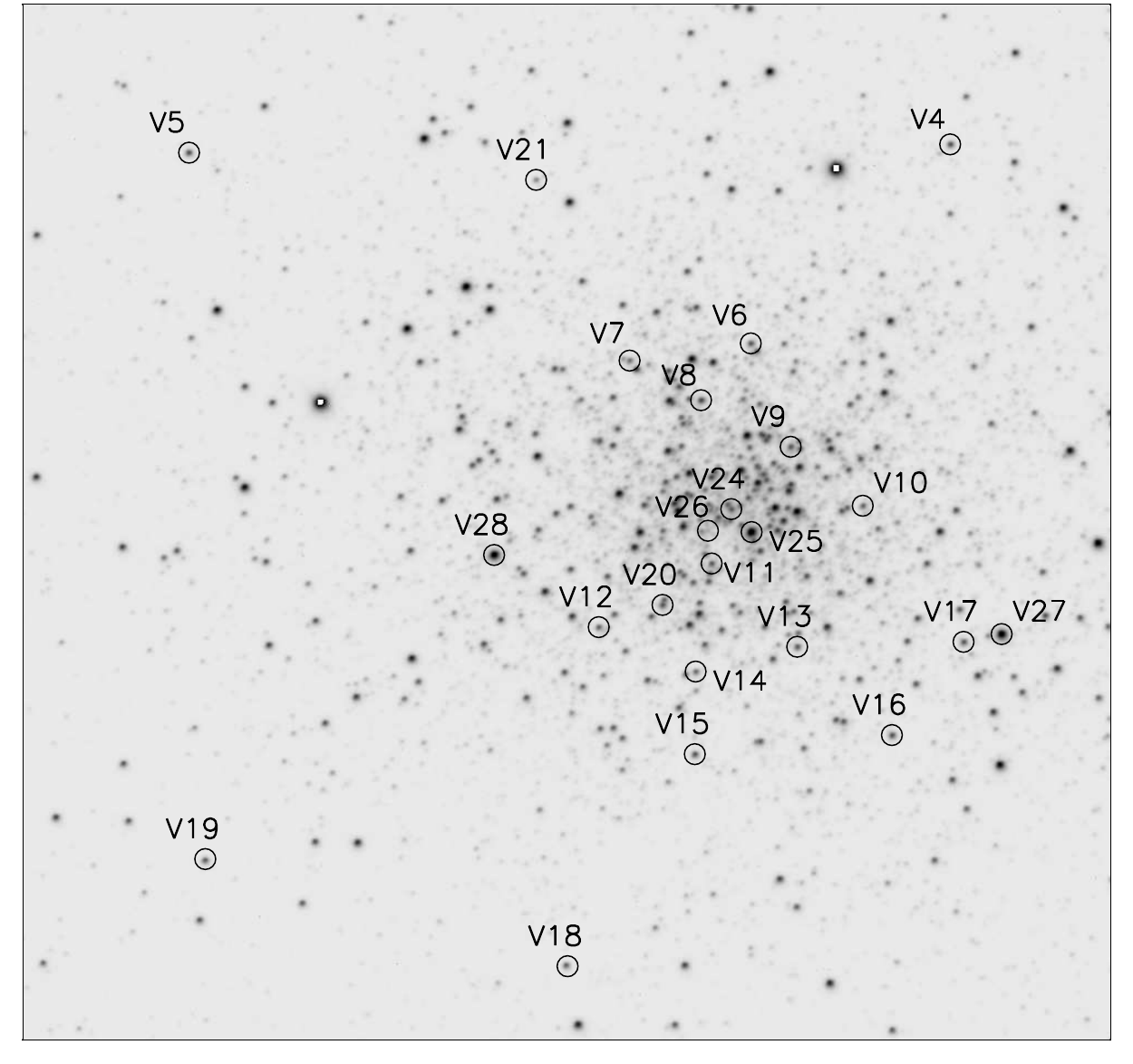}
\caption{Finding chart built from the $V$ reference image. The FoV is approximately 4.8 $\times$ 4.8 arcmin$^2$.
Note that in the chart North is down and East is to the left.}
\label{chart6171}
\end{center}
\end{figure*}

\vspace{17.0cm}
\bibliographystyle{Wiley-ASNA}
\bibliography{6171_biblio}
\end{document}